\documentclass[
showkeys,12pt,
preprint,preprintnumbers,nofootinbib,
groupedaddress,superscriptaddress,amsmath,amssymb]{revtex4}
\usepackage{graphicx}
\usepackage{dcolumn}
\usepackage{bm}
\usepackage{amssymb}
\usepackage{amsmath}
\usepackage{epsfig}    
\usepackage{color}
\usepackage{slashed}
\usepackage{hhline}

\def\be{\begin{equation}}
\def\ee{\end{equation}}
\newcommand{\bea}{\begin{eqnarray}}
\newcommand{\eea}{\end{eqnarray}}
\newcommand{\nn}{\nonumber}

\numberwithin{equation}{section}

\begin{document}

 \begin{flushright} {KIAS-P20034}, APCTP Pre2020-012  \end{flushright}

\title{Axion and neutrino mass from a hidden gauge symmetry model}

\author{Takaaki Nomura}
\email{nomura@kias.re.kr}
\affiliation{School of Physics, KIAS, Seoul 02455, Republic of Korea}

\author{Hiroshi Okada}
\email{hiroshi.okada@apctp.org}
\affiliation{Asia Pacific Center for Theoretical Physics, Pohang 37673, Republic of Korea}
\affiliation{Department of Physics, Pohang University of Science and Technology, Pohang 37673, Republic of Korea}

\author{Seokhoon Yun}
\email{SeokhoonYun@kias.re.kr}
\affiliation{School of Physics, KIAS, Seoul 02455, Republic of Korea}

\date{\today}

\begin{abstract}
We  propose a model of dark sector described by gauged hidden $U(1)_H$ symmetry
in which neutrino masses are generated at one-loop level and axion is induced by assigning Peccei-Quinn charge to fermions in dark sector relevantly.
Then our scenario connects exotic fermion mass generation, neutrino mass matrix and axion through scalar fields associated with $U(1)_H$ and Peccei-Quinn symmetry breaking.
We investigate neutrino mass formula, lepton flavor violation, anomalous magnetic moment of muon, dark matter relic density and axion couplings, 
which are originated from interactions among our dark sector and the standard model particles.
 \end{abstract}
\maketitle
\section{Introduction}

There are several issues which are unsolved within the standard model (SM) of particle physics such as 
an existence of dark matter (DM), non-zero neutrino masses and mixings and the strong CP problem. 
We thus expect such issues are explained by the physics beyond the SM whose description is an open question.
One of the plausible scenario is that new physics sector is described by hidden gauge symmetry under which only new fields are charged 
as the SM part is also described by gauge symmetry; here we denote such new physics sector as dark sector controlled by hidden gauge symmetry.
Interestingly,  by applying such hidden gauge symmetry, we can stabilize DM candidate and forbid neutrino mass generation at tree level to explain smallness of its value~\cite{Nomura:2018kdi, Cai:2018upp, Nomura:2018ibs, Nomura:2017wxf, Ko:2017uyb, Ko:2016uft, Ko:2016wce, Ko:2016ala, Ko:2014loa, Ko:2014eqa,Ma:2013yga,Yu:2016lof,Ko:2016sxg,Nomura:2020azp}.
Then neutrino mass can be generated at loop level by introducing appropriate interactions among dark sector fields and leptons.
In addition, fermions in dark sector can be chiral under gauge symmetry obtaining their mass after spontaneous symmetry breaking, 
and it is possible to assign Peccei-Quinn (PQ) charges~\cite{Peccei:1977hh,Peccei:1977ur} consistently inducing axion~\cite{Weinberg:1977ma,Wilczek:1977pj} to explain the strong CP problem.

In this paper, we construct a model of dark sector described by hidden $U(1)_H$ gauge symmetry.
In this model new fermions are chiral under $U(1)_H$ and PQ charges are assigned consistently; gauge anomaly cancellation condition requires exotic leptons and quarks.
We also introduce several scalar fields in dark sector to realize neutrino mass generation at one-loop level and to give DM candidate stabilized by remnant $Z_2$ symmetry originated from $U(1)_H$.
Remarkably our scenario connects exotic fermion masses, neutrino mass matrix and axion, by scalar fields associated with $U(1)_H$ and PQ symmetry breaking~\footnote{Models relating axion and radiative neutrino mass generation are discussed, for examples, in refs.~\cite{Ma:2017zyb, Ma:2017vdv, Suematsu:2017kcu}}.  
Note also that our model provides multi-component DM scenario since both the lightest $Z_2$ odd particle and axion can be good DM candidates.
In addition, interactions among dark sector particles and the SM leptons induce lepton flavor violating(LFV) processes and contribution to muon anomalous magnetic moment (muon $g-2$).
 
After constructing framework of the model, we analyze neutrino mass matrix induced by one-loop diagram, LFV processes, muon $g-2$, relic density of DM and axion couplings. 
Then consistency of the model is discussed by considering constraints from observed data.
We then provide benchmark point in the model. 

This paper is organized as follows.
In Sec.~II, we show our model, 
and show mass formula for particles in dark sector. 
In Sec.~III we consider phenomenology from dark sector deriving the analytical forms of neutrino mass matrix, LFVs, muon anomalous magnetic moment and relic  density of DM, and axion couplings. 
We then conclude and discuss in Sec.~IV.


\begin{center} 
\begin{table}
\begin{tabular}{|c||c|c|c|c||c|c|c|c|c|}\hline\hline  
&\multicolumn{4}{c||}{Fermions} & \multicolumn{4}{c|}{Bosons} \\\hline
Fermions&~ $E_{L/R}^\alpha$ ~&~ $N_{L/R}^\alpha$ ~&~ $U_{L/R}^\alpha$ ~&~ $D^\alpha_{L/R}$ ~&~
$\Phi$ ~&~ $X$ ~&~ $\eta$ ~&~ $\varphi$ ~
\\\hline 
$SU(3)_C$ & $\bm{1}$  & $\bm{1}$&
 $\bm{3}$  & $\bm{3}$    & $\bm{1}$   & $\bm{1}$  & $\bm{1} $  & $\bm{1} $
   \\\hline 
 $SU(2)_L$ & $\bm{1}$  & $\bm{1}$  & $\bm{1}$  &
   $\bm{1}$   & $\bm{1}$  & $\bm{1}$   & $\bm{2}$ & $\bm{1}$   
  \\\hline 
$U(1)_Y$ & $-1$ & $0$  & $\frac{2}{3}$  & $-\frac13$  & $0$  &   $0$ &  $\frac12$  &  $0$ 
 \\\hline
 $U(1)_{H}$ & $1/3$ & $-1/-3$  & $-1/-3$   & $1/3$ & $2$ & $1$  & $-3$  & $-2$
 \\\hline
  $U(1)_{PQ}$ & $-1/-3$ & $1/3$  & $1/-1$   &  $-1/-3$ & $2$ & $-1$  & $3$  & $2$
 \\\hline
\end{tabular}
\caption{Field contents of fermions
and their charge assignments under $SU(3)_C\times SU(2)_L\times U(1)_Y\times U(1)_H \times U(1)_{PQ}$, where each of the flavor index is defined as $\alpha\equiv 1-3$.}
\label{tab:fb}
\end{table}
\end{center}

\section{ Model construction }
In this section, we construct our model of dark sector with $U(1)_H$ gauge symmetry and show mass spectrum.
As for fermion sector, we introduce three families of isospin singlet exotic leptons and quarks; $E$, $N$, $U$, $D$, which have nonzero charges under hidden local $U(1)_H$ symmetry, where the SM fields are all neutral under this additional symmetry.
Note that exotic fermions are chiral under $U(1)_H$ charge assignment and massless before spontaneous symmetry breaking.
In addition, we assign Peccei-Quinn(PQ) charge to these fermions.
As for scalar sector, we introduce inert scalar fields $X$ and $\eta$ , each of which are singlet and doublet under $SU(2)_L$,
while $\Phi$ and $\varphi$ have nonzero vacuum expectation values (VEVs), both of which are singlet under $SU(2)_L$ and their VEVs are denoted by $v_\Phi/\sqrt2$ and $v_\varphi/\sqrt2$, respectively. Here, the SM Higgs is defined by  $H$ and its VEV is characterized by $v_H/\sqrt2$.
All the new field contents and their assignments are represented  in Table~\ref{tab:fb}.
Then anomaly cancellations of $U(1)_H$ are found by computing the following four relations for each generation:
\[ [U(1)_Y]^2U(1)_{H}, \ [U(1)_{H}]^2U(1)_Y, \ [U(1)_{H}]^3, \  U(1)_{H}. \]
%
The renormalizable new terms in Yukawa Lagrangians are found to be 
\begin{align}
-{\cal L}&=
y^E_{\alpha\alpha}\bar E^\alpha_L E^\alpha_R \varphi+ y^N_{\alpha\alpha}\bar N^\alpha_L N^\alpha_R \varphi^*
+ y^U_{\alpha\alpha}\bar U^\alpha_L U^\alpha_R \Phi +  y^D_{\alpha\alpha}\bar D^\alpha_L D^\alpha_R \varphi
+ y'^N_{\alpha\beta}\bar N^{c,\alpha}_L N^\beta_L \varphi^* \nn\\
& + y^{Ee}_{\alpha i}\bar E^{\alpha}_L e^i_R X  + y^{Uu}_{\alpha i}\bar U^{\alpha}_L u^i_R X^*  
+y_\eta^{i\alpha} \bar L_{L}^i\tilde \eta N_R^\alpha
+y_{\alpha i}^{Dd} \bar D_{L}^i d_{R_a} X+{\rm h.c.}, 
 \label{eq:lag-ykw}
\end{align}
where $\tilde \eta\equiv \sigma_2 \eta^*$, $\sigma_2$ being second Pauli matrix, $i=1-3$ and $(\alpha,\beta)=1-3$ are indices of flavors of SM and exotic fermions, respectively. 
Notice here that the first four terms are diagonal without loss of generality.
The terms in first line of the right-hand side will provide exotic fermion mass terms 
after the spontaneous symmetry breaking as we will see below.

Scalar potential in our model is given by
\begin{align}
V&=
\mu_\varphi^2 |\varphi|^2 +\mu_\Phi^2 |\Phi|^2 +\mu^2_{X} |X|^2 + \mu_H^2 |H|^2 + \mu_\eta^2 |\eta|^2 
+\mu_0\left[X^2\varphi +{\rm h.c.}  \right] +\lambda_0[\eta^\dag HX^*\Phi+{\rm h.c.}]\nn\\
&+\lambda_\varphi|\varphi|^4  +\lambda_\Phi |\Phi|^4 +\lambda_X |X|^4+ \lambda_{H} |H|^4+ \lambda_{\eta} |\eta|^4
+\lambda_{\varphi\Phi}|\varphi|^2  |\Phi|^2+\lambda_{\varphi X}|\varphi|^2  |X|^2\nn\\
&+\lambda_{\varphi H}|\varphi|^2  |H|^2+\lambda_{\varphi\eta}|\varphi|^2  |\eta|^2
+\lambda_{\Phi X} |\Phi|^2 |\Phi|^2+\lambda_{\Phi H} |\Phi|^2 |H|^2+\lambda_{\Phi \eta} |\Phi|^2 |\eta|^2\nn\\
&+\lambda_{X H} |X|^2 |H|^2 +\lambda_{X \eta} |X|^2 |\eta|^2 +\lambda_{\eta H} |H|^2|\eta|^2
+\lambda'_{H\eta} |H^\dag\eta|^2,
\label{eq:lag-pot}
\end{align}
where the scalar fields are parameterized as 
\begin{align}
&H =\left[
\begin{array}{c}
w^+\\
\frac{v_{H}+h+iz}{\sqrt2}
\end{array}\right],\ 
\eta =\left[
\begin{array}{c}
\eta^+\\
\frac{\eta_R+i\eta_I}{\sqrt2}
\end{array}\right],\ 
\varphi=\frac{v_\varphi+\varphi_R+iz_\varphi}{\sqrt2},\
\Phi=\frac{v_\Phi+\phi_R+iz_\Phi}{\sqrt2},\
X=\frac{\chi_R+i\chi_I}{\sqrt2},
\label{component}
\end{align}
where $w^\pm$, and $z$ are respectively absorbed by the longitudinal degrees of freedom of charged SM gauge boson $W^\pm$ and neutral SM gauge boson $Z$. 
A linear combination of $z_\varphi$ and $z_\Phi$ is also eaten by neutral $U(1)_H$  gauge boson $Z'$,
while another combination is identified to be axion.
{Our $Z'$ boson mass is given by VEVs of $\varphi$ and $\Phi$, and it should be larger than $\sim 10^8$ GeV by constraint from axion decay constant as we discuss below.
Thus $Z'$ tends to be much heavier than TeV scale and we do not discuss phenomenology associated with it.
Also mass scale associated with $\varphi$ and $\Phi$ should be much higher than electroweak scale and mixing among neutral CP-even components of them and SM Higgs boson  
will be negligible. We thus do not consider these CP-even scalar components and assume the SM Higgs boson comes from $H$ just as in the SM case.
Notice that $U(1)_H$ breaks into remnant $Z_2$ symmetry due to our charge assignment where $U$, $D$, $E$, $N$, $X$ and $\eta$ are $Z_2$ odd and the other fields are even.
As a result the lightest $Z_2$ odd particle is stable and can be good DM candidate if it is neutral.}

{
{\it Inert scalar masses}:
In the CP even inert bosons in basis of  $(\chi_R,\eta_R)$, we can write the mass term such that
\begin{equation}
 \mathcal{L}_M^{\chi_R \eta_R}  
 = \frac12 \begin{pmatrix} \chi_R \\ \eta_R \end{pmatrix}^T
\begin{pmatrix} \mu_\chi^2 + \frac{\lambda_{\varphi X} v_\varphi^2 + \lambda_{\Phi X} v_\Phi^2 + \lambda_{XH} v_H^2}{2}  + \sqrt{2} \mu_0 v_\varphi & \frac{\lambda_0}{2} v_\Phi v_H \\
\frac{\lambda_0}{2} v_\Phi v_H  & \mu_\eta^2 + \frac{\lambda_{\varphi \eta} v_\varphi^2 + \lambda_{\Phi \eta} v_\Phi^2 + (\lambda_{\eta H}+\lambda'_{\eta H}) v_H^2}{2} \end{pmatrix}
\begin{pmatrix} \chi_R \\ \eta_R \end{pmatrix},
\end{equation}
where mixing between $\chi_R$ and $\eta_R$ is induced by 7th term of the potential Eq.~\eqref{eq:lag-pot}.
Similarly we obtain mass matrix for $\chi_I$ and $\eta_I$ as 
\begin{equation}
 \mathcal{L}_M^{\chi_I \eta_I}  
 = \frac12 \begin{pmatrix} \chi_I \\ \eta_I \end{pmatrix}^T
\begin{pmatrix} \mu_\chi^2 + \frac{\lambda_{\varphi X} v_\varphi^2 + \lambda_{\Phi X} v_\Phi^2 + \lambda_{XH} v_H^2}{2}  - \sqrt{2} \mu_0 v_\varphi & -\frac{\lambda_0}{2} v_\Phi v_H \\
-\frac{\lambda_0}{2} v_\Phi v_H  & \mu_\eta^2 + \frac{\lambda_{\varphi \eta} v_\varphi^2 + \lambda_{\Phi \eta} v_\Phi^2 + (\lambda_{\eta H}+\lambda'_{\eta H}) v_H^2}{2} \end{pmatrix}
\begin{pmatrix} \chi_I \\ \eta_I \end{pmatrix}.
\end{equation} }
After diagonalizing the mass matrices, we define the mixing and its mass eigenvalue as follows:
\begin{align}
&\chi_R = s_{\theta_R} H_1 + c_{\theta_R} H_2,\quad \eta_R =-c_{\theta_R} H_1 + s_{\theta_R} H_2,\quad 
{\sin {2\theta_R}=\frac{\lambda_0 v_\Phi v_H}{m_{H_2}^2 - m_{H_1}^2},}\\
&\chi_I = s_{\theta_I} A_1 + c_{\theta_I} A_2,\quad \eta_I =-c_{\theta_I} A_1 + s_{\theta_I} A_2,\quad 
{\sin {2\theta_I}= -\frac{\lambda_0 v_\Phi v_H}{m_{A_2}^2 - m_{A_1}^2},}
\end{align}
where $s_{\theta_{R/I}}(c_{\theta_{R/I}})$ is the short-hand symbol of $\sin{\theta_{R/I}}(\cos{\theta_{R/I}})$ 
and mass eigenvalues of $\{H_1, H_2, A_1, A_2\}$ are denoted as $\{ m_{H_1}, m_{H_2}, m_{A_1}, m_{A_2} \}$ correspondingly.


{
{\it Exotic charged fermion masses}:
We obtain masses of exotic charged fermions after spontaneous symmetry breaking from Yukawa interactions in Eq.~\eqref{eq:lag-ykw}.
Then masses of $\{E, D, U\}$ are given by
\begin{equation}
m_E = \frac{1}{\sqrt{2}} y^E v_\varphi,  \quad m_D = \frac{1}{\sqrt{2}} y^D v_\varphi, \quad m_U = \frac{1}{\sqrt{2}} y^E v_\Phi, 
\end{equation}
where we omit flavor index.}

{\it  Exotic neutral fermion masses}:
Here we discuss the heavier neutral fermion sector in the following.
We have a mass matrix of neutral fermion in basis of  $\Psi\equiv [N^C_R,N_L]^T$,
and they are given by six by six matrix as
\begin{align}
M_{\Psi}
&\equiv
\left[\begin{array}{cc}
{\bf 0}_{3\times3} & (M_N^\dag)_{3\times3}  \\ 
(M_N^*)_{3\times3} & (M'_{N})_{3\times3} \\ 
\end{array}\right],
 \end{align}
where the elements are defined by $M_N \equiv y^N v_\varphi/\sqrt{2}$ and $M'_N \equiv y'^N v_\varphi/\sqrt{2}$.
Then the mass eigenstate and its mixing is respectively defined by $D_{\psi}=V M_{\Psi} V^T$, and 
\begin{align} 
\left[\begin{array}{c}
N^C_R  \\ 
N_L \\ 
\end{array}\right]_i
=(V^T)_{ij} \psi_j , \  { i,j=1\sim6},
\end{align}
where $V$ is  the  unitary mixing matrix with six by six, and $\psi_i$ is the mass eigenstate, and $D_{\psi}$ is mass eigenvalue.

\begin{figure}[t]
\begin{center}
\includegraphics[width=90mm]{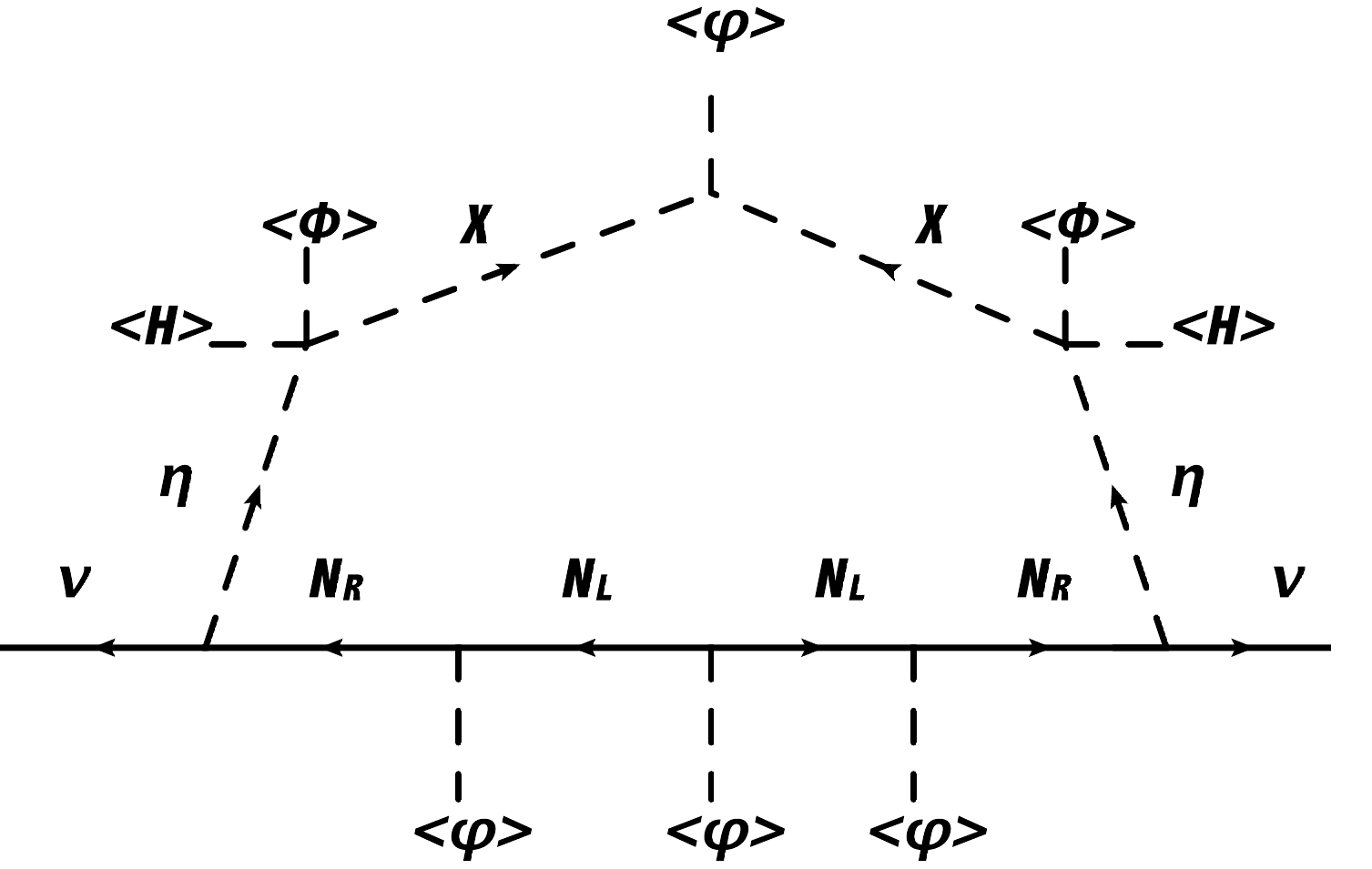} 
\caption{The one loop diagram to generate neutrino masses. } 
 \label{fig:neut}
\end{center}\end{figure}

\section{Phenomenology from dark sector}

In this section, we discuss phenomenology originated from $U(1)_H$ dark sector such as neutrino mass generation, LFV processes, muon $g-2$ and axion.

\subsection{Neutrino mass generation}

In this model active neutrino mass matrix is generated at one loop level after spontaneous symmetry breaking.
The relevant one-loop diagram is shown in fig.~\ref{fig:neut}, and the analytic form
 is given by
\begin{align}
&(m_\nu)_{ij}\approx 
\frac{2}{(4\pi)^2} \sum_{a=1}^6
{Y_{ia} D_{\psi_a} Y^T_{aj}}
(3M_1^2 F_I^a - M_2^4 F_{II}^a +M^6_3 F_{III}^a),\\
&M_1^2\equiv \frac12[m_{H_1}^2+m_{H_2}^2-m_{A_1}^2-m_{A_2}^2+(m_{H_1}^2-m_{H_2}^2)c_{2R}-(m_{A_1}^2-m_{A_2}^2)c_{2I}],\nn\\
&M_2^4\equiv \frac12[-2m_{H_1}^2m_{H_2}^2+2m_{A_1}^2m_{A_2}^2-(m_{H_1}^2-m_{H_2}^2)(m_{A_1}^2+m_{A_2}^2)c_{2R}+(m_{H_1}^2+m_{H_2}^2)(m_{A_1}^2-m_{A_2}^2)c_{2I}],\nn\\
&M_3^6\equiv  m_{H_1}^2m_{H_2}^2m_{A_2}^2 -m_{A_1}^2[m_{A_2}^2m_{H_2}^2c_R^2+m_{H_1}^2(-m^2_{H_2}s_I^2+m_{A_2}^2 s_R^2)],\nn\\
&F_I^a=\int\frac{dx_1dx_2dx_3dx_4dx_5\delta(1-x_1-x_2-x_3-x_4 - x_5)}{(x_1 D_{\psi_a}^2+x_2 m_{H_1}^2+x_3 m_{H_2}^2
+x_4 m_{A_1}^2+x_5 m_{A_2}^2)^I },\ {\rm I=1,2,3},
\end{align}
where $Y_{ia}\equiv \sum_{\rho=1}^3 y_{\eta}^{i\rho}V^\dag_{\rho a}$.
Since one diagonalizes neutrino mass matrix as $D_\nu\approx V_{MNS}^T m_\nu V_{MNS}$, 
we can rewrite Yukawa coupling in terms of neutrino oscillation data and some parameters as~\cite{Casas:2001sr}:
\begin{align}
(y_\eta)_{ij} = (V^*_{MNS})_{ia} \sqrt{D_\nu} V O (T^T)^{-1},
\end{align}
where $T$ is an upper-right triangle matrix that comes from 
$R_{ij}  \equiv 2 V_{ia} D_{\psi_a}(3M_1^2 F_I^a - M_2^4 F_{II}^a +M^6_3 F_{III}^a)(V^T)_{aj} /(4\pi)^2 = T^TT$~\footnote{This decomposition can be generally possible when $R$ is a symmetric matrix.}, $V_{MNS}$ is Maki-Nakagawa-Sakata mixing matrix~\cite{Gonzalez-Garcia:2014bfa}, and $O$ is an arbitral three by three matrix that satisfies $OO^T=O^TO=1_{3\times3}$.
Satisfying the neutrino oscillation data is rather easy task due to $O$, and all we should take care is
to consider the constraints of lepton flavor violations that will be discussed in the next subsection. 

\subsection{Muon $g-2$ and LFVs}

 {\it Muon $g-2$}: {The muon anomalous magnetic moment($\Delta a_\mu$) has been observed in the E821 experiment at Brookhaven National Lab (BNL)~\cite{Bennett:2006fi}
 and its discrepancy from the SM prediction is estimated as~\cite{PDG}
\begin{align}
\Delta a_\mu=(26.1\pm 7.9)\times10^{-10}, \label{eq:exp-g2mu}
\end{align}
where it indicates $3.3 \sigma$ deviation. 
A $3.7\sigma$ deviation was recently obtained by the lattice calculations as $\delta a_\mu = (27.4\pm 7.3)\times 10^{-10}$~\cite{Blum:2018mom} and $\delta a_\mu = (27.06\pm 7.26)\times 10^{-10}$~\cite{Keshavarzi:2018mgv}.}
Our $\Delta a_\mu$ is induced at one-loop level via the Yukawa interactions associated with $y^{eE}$ and $y_\eta$ where the $H_{1,2}$, $A_{1,2}$ and $E$ propagate inside the loop diagram.
The analytic form is computed as
\begin{align}
&\Delta a_{\mu}\approx -\frac{2m_\mu^2}{(4\pi)^2} 
\left(\sum_{a=1-6}Y_{2a}Y^\dag_{a2} G_{II}(\psi_a,\eta^\pm)\right. \\
&\left.-
\sum_{\alpha=1-3}y^{Ee\dag}_{2\alpha}y^{Ee}_{\alpha 2}
\left[s_R^2G_{II}(E_\alpha,H_1)+c_R^2G_{II}(E_\alpha,H_2)+s_I^2G_{II}(E_\alpha,A_1) + c_I^2G_{II}(E_\alpha,A_2)\right]\right),\nn\\
&G_{II}(a,b)\approx\frac{2 m_a^6+3m_a^4m_b^2-6m_a^2m_b^4+m_b^6+12m_a^4m_b^2\ln\left[\frac{m_b}{m_a}\right]}{12(m_a^2-m_b^2)^4},
\end{align}
where $m_a\neq m_b$ are assumed in $G_{II}$.

{\it Lepton flavor violations (LFVs)}: LFV processes of $\ell \to \ell' \gamma$ are arisen from the same term as the $(g-2)_\mu$, and their forms are given by
\begin{align}
&BR(\ell_i\to \ell_j \gamma)
\approx\frac{48\pi^3C_{ab} \alpha_{em}}{(4\pi)^4G_F^2}
\left| 
\sum_{a=1-6}Y_{ja}Y^\dag_{ai} G_{II}(\psi_a,\eta^\pm)\right. \\
&\left.-
\sum_{\alpha=1-3}y^{Ee\dag}_{j\alpha}y^{Ee}_{\alpha i}\left[s_R^2G_{II}(E_\alpha,H_1)+c_R^2G_{II}(E_\alpha,H_2)+s_I^2G_{II}(E_\alpha,A_1) + c_I^2G_{II}(E_\alpha,A_2)\right] \right|^2,\nn
\end{align}
where $\alpha_{em}\approx1/137$ is the fine-structure constant, $G_F\approx1.17\times10^{-5}$ GeV$^{-2}$ is the Fermi constant,
and $C_{21}\approx1$, $C_{31}\approx 0.1784$, $C_{32}\approx0.1736$. 
Experimental upper bounds are given by~\cite{TheMEG:2016wtm, Adam:2013mnn}: 
\begin{equation}
{\rm BR}(\mu\to e \gamma)\lesssim 4.2\times 10^{-13},\ 
{\rm BR}(\tau\to e \gamma)\lesssim 3.3\times 10^{-8},\ 
{\rm BR}(\tau\to \mu \gamma)\lesssim 4.4\times 10^{-13},
\end{equation}
where we define $\ell_1\equiv e$,  $\ell_2\equiv \mu$, and  $\ell_3\equiv \tau$. 


\subsection{Axion} 
 
Due to the remaining global PQ symmetry in the model, there is the Nambu-Goldstone boson after the spontaneous breaking by the non-zero VEV of $\Phi$ and $\varphi$.
As already mentioned, this corresponding pseudo-scalar boson is called `axion'.

Note that the PQ symmetry is anomalous for $SU(3)_C$ and $U(1)_Y$ leading to the presence of the anomalous axion coupling to the gluon and photon
\bea
\frac{g_s^2}{32\pi^2}\frac{a}{f_a}G\tilde{G} + n_\gamma\frac{e^2}{32\pi^2}\frac{a}{f_a}F\tilde{F} \, ,
\eea
where $g_s$ and $e$ corresponds to the coupling constant of  $SU(3)_C$ and $U(1)_{\rm EM}$ gauge, respectively, and $f_a$ denotes the axion decay constant.
Here the anomalous gluon coupling is normalized by redefinition of $f_a$ then we find
\bea
f_a & = & \frac{v_\varphi v_\Phi}{\sqrt{v_\varphi^2 + v_\Phi^2}}, \\  
n_\gamma & = &  8/3 \, .
\eea 
At scales below the strong confinement $\Lambda_{\rm QCD}$, this symmetry breaking term of the anomalous gluon coupling contributes to the axion potential then one could solve the naturalness problem of $\theta_{\rm QCD} < 10^{-10}$~\cite{Baker:2006ts} dynamically.

Considering that the exotic fermions are heavy comparable to the $v_\Phi$ or $v_\varphi$, the phenomenologically important axion couplings are the anomalous couplings to the vector boson which is categorized as the `KSVZ'~\cite{Kim:1979if,Shifman:1979if} axion model.
The most stringent constraints on $f_a$ is derived from the observed neutrino signal of the SN 1987A~\cite{Raffelt:1996wa} given as~\cite{Raffelt:2006cw,Fischer:2016cyd}
\bea
f_a \geq 4\times 10^{8}\,{\rm GeV} \, .
\eea

There are also the couplings to the $U(1)_H$ gauge boson (so called `dark axion portal'~\cite{Kaneta:2016wvf,Kaneta:2017wfh} terms).
However, these couplings are phenomenologically less important so we will not discuss in this paper.

\subsection{Dark matter}
{In our model we have two DM candidates. 
One is the lightest neutral exotic particle which is odd under remnant $Z_2$ symmetry from local $U(1)_H$.
The other one is axion which comes from $\Phi$ and $\varphi$ as we discussed above.
Here, we assume the DM candidate of first type as $H_2$ that is nearly identified to be $\chi_R$ considering small $\theta_R$. 
In the following we denote $X_D \equiv H_2$ with mass $m_{X_D}$.}

We have several relevant interactions to explain relic density; 
inducing annihilation processes $X_D X_D \to\ell_i\bar\ell_j/u_i\bar u_j/d_i\bar d_j/2H_{SM}$,
where the first three modes come from $y^{Ee},y^{Uu},y^{Dd}$, and the last mode arises from Higgs potential.
However since all the modes except $y^{Ee}$ is restricted by the constraints of direct detection searches whose couplings are of the order $10^{-3}$. Thus, 
the dominant cross section to explain the relic density is induced from $y^{Ee}$ interaction and
given in terms of relative velocity and found d-wave dominant~\cite{Chiang:2017zkh};  
\begin{align}
\sigma v_{rel}(2X_D \to \ell_i\bar\ell_j)
\approx \frac{v^4_{rel}}{240\pi m_{X_D}^2}
\left|
\sum_{\alpha=1}^3\frac{y^{Ee}_{\alpha i} y^{Ee\dag}_{j\alpha}}{\left(1+\frac{m^2_{E_\alpha}}{m_{X_D}^2}\right)^2}\right|^2,
\end{align}
where we ignored charged lepton masses.
The relic density of DM can be estimated by solution of Boltzmann equation and 
we obtain relevant cross section to explain observed relic density $\Omega h^2 \sim 0.12$~\cite{PDG}.

\begin{figure}[t]
\begin{center}
\includegraphics[width=90mm]{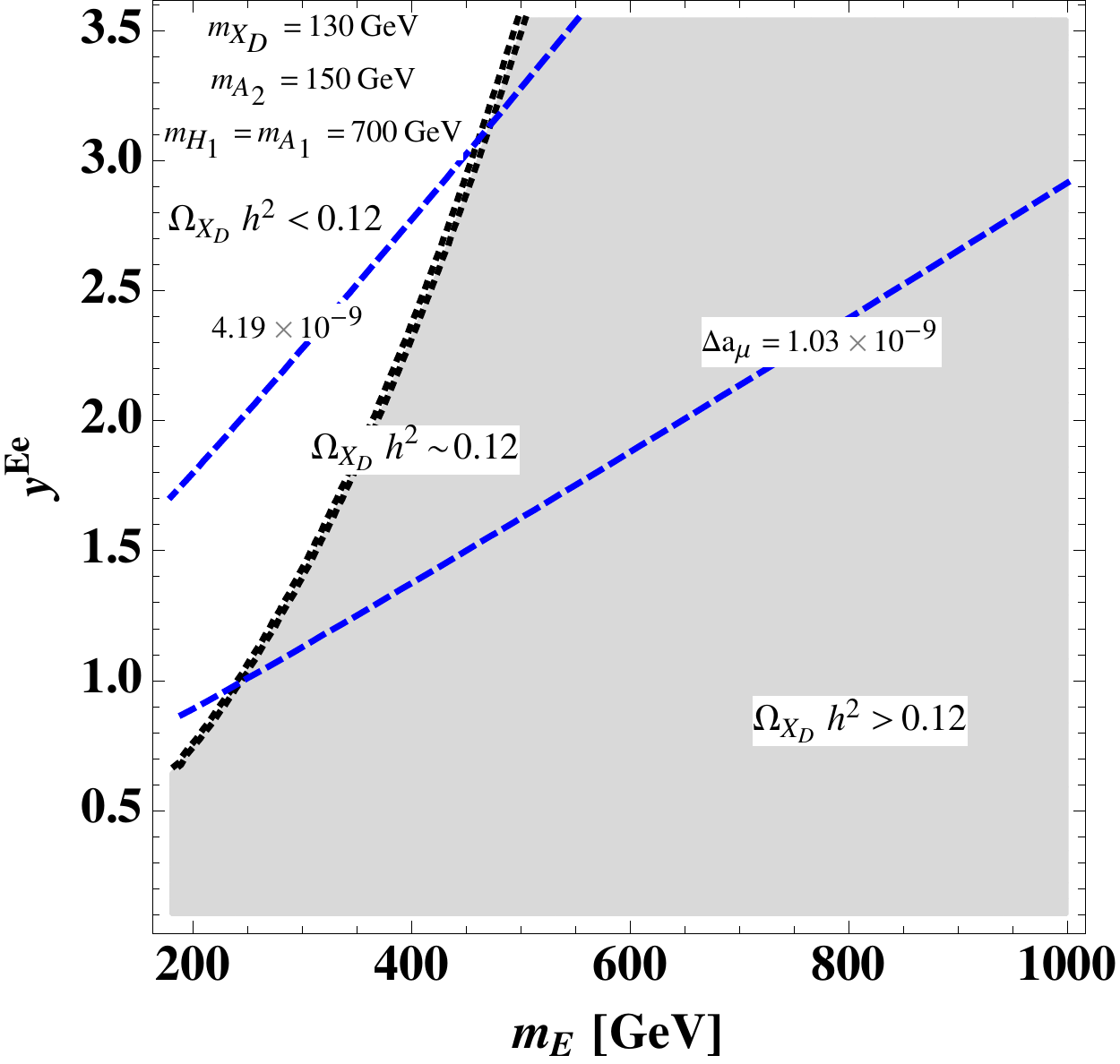} 
\caption{Contour plot for $X_D$ DM relic density and muon $g-2$ where $y^{Ee} \equiv y^{Ee}_{12}=y^{Ee}_{22}=y^{Ee}_{32}$, $m_E \equiv M_{E_1} =  M_{E_2} = M_{E_3}$ and 
 inert scalar boson masses are indicated on the figure.} 
 \label{fig:DM}
\end{center}\end{figure}

Then, the range of cross section to explain the correct relic density at 2 $\sigma$ C.L. is estimated as 
 \begin{align}
1.776\times 10^{-9}\ {\rm GeV} \lesssim \sigma v_{rel}\lesssim 1.9697\times 10^{-9}\ {\rm GeV},
\end{align}
where $v_{rel}\approx0.3$ is used.
For illustration, we show contour plot for $X_D$ DM relic density and $\Delta a_\mu$ in Fig.~\ref{fig:DM} in the plane of $y^{Ee} \equiv y^{Ee}_{12}=y^{Ee}_{22}=y^{Ee}_{32}$ and $M_E \equiv M_{E_1} =  M_{E_2} = M_{E_3}$
fixing other parameters as $s_R = s_I = 0.1$, $m_{X_D} =130$ GeV, $m_{A_2} = 150$ GeV and $m_{H_1}=m_{A_2} = 700$ GeV.
The region between blue dashed lines can explain muon $g-2$ within 2 $\sigma$.
{We find that the relic density is overabundant for small $y_E$ and heavy $M_E$ region.
For region with $\Omega_{X_D} h^2 < 0.12$ of $X_D$, we expect axion DM can compensate the lack of relic density. }

{
{\it Benchmark point;}
Instead of showing global analysis, we demonstrate a benchmark point satisfying neutrino data, flavor constraints, relic density of DM by $X_D$, and muon $g-2$.
The particle masses and observables are given by
\begin{align}
& \{m_{H_1}, m_{A_1}, m_{A_2} \} \simeq \{24600,2210 ,179 \} {\rm [GeV]}, \  \{ m_{E_1}, m_{E_1}, m_{E_2} \} \simeq \{156,162 ,174 \} {\rm [GeV]}, \nonumber \\
&M_X\approx 125{\rm GeV},\ \Delta a_\mu\approx 3.40\times10^{-9}, \sigma v_{\rm rel}\approx 1.93\times 10^{-9}{\rm GeV^{-2}}, \nonumber \\
& {\rm BR}(\mu\to e\gamma)\approx 4.50\times 10^{-17},\ 
 {\rm BR}(\tau\to e\gamma)\approx 2.88\times 10^{-18},\  {\rm BR}(\tau\to \mu\gamma)\approx 4.72\times 10^{-18},
\end{align}
where we have taken range of input parameters as $[10^2-10^5]$ GeV for masses, [$0.01-\sqrt{4\pi}$] for dimensionless parameters, and $[(-0.1)$ -- $0.1]$ for $s_{R,I}$ in searching for the benchmark point.
Here mass of exotic quarks are omitted since they are irrelevant for neutrino mass, DM and LFV in our scenario.}

We would like to note that the axion can constitute the cold dark matter adequately through the misalignment mechanism~\cite{Preskill:1982cy,Abbott:1982af,Dine:1982ah}, i.e. the coherent oscillation of the axion field.
If the PQ symmetry is broken before and during inflation, the relic abundance of the axion cold dark matter is given by~\cite{Borsanyi:2016ksw,Bae:2008ue,Wantz:2009it,Ballesteros:2016xej}
\bea
\Omega_{a}h^2\approx 0.12 \left(\frac{f_A}{9\times 10^{11}\,{\rm GeV}}\right)^{1.165} F\theta_i^2 \, ,
\eea
where $F$ is the anharmonic effect due to an its periodic potential and $\theta_i = a_i/f_a$ is the initial misalignment angle.

\section{ Conclusions and discussions}
We have proposed a model in which dark sector is described by hidden $U(1)_H$ gauge symmetry, and exotic fermions are chiral under $U(1)_H$ obtaining masses via spontaneous symmetry breaking.
In our charge assignment $U(1)_H$ is broken to remnant $Z_2$ symmetry and the lightest $Z_2$ odd particle is stable being good DM candidate if it is neutral.
Introducing several scalar fileds with non-zero $U(1)_H$ charge, we can generate neutrino mass via one-loop diagram in which particles in dark sector propagate.
Furthermore we can assign relevant PQ charges to dark sector fermions and 'KSVZ' type axion can be obtained from scalar fields whose VEVs break $U(1)_H$ and PQ symmetry.
Interestingly these scalar fields play roles of giving exotic fermion masses, realizing active neutrino mass and providing axion to solve strong CP problem.
Note also that our model is multi-component DM scenario since axion can be also candidate of DM in addition to $Z_2$ odd particle.

We have then analyzed neutrino mass matrix, lepton flavor violating processes, muon $g-2$, DM relic density and axion couplings.
Our neutrino mass matrix can accommodate  with observed data by appropriately choosing Yukawa couplings among dark sector particle and the SM leptons.
Also muon $g-2$ can be explained by Yukawa interactions among muon and dark sector particles which can also induce DM annihilation processes consistent with observed relic density of DM.
In addition axion couplings to gluon and photon are derived where we find clear relation between them and constraint of axion decay constant is given.


\section*{Acknowledgments}
\vspace{0.5cm}
{\it
This research was supported by an appointment to the JRG Program at the APCTP through the Science and Technology Promotion Fund and Lottery Fund of the Korean Government. This was also supported by the Korean Local Governments - Gyeongsangbuk-do Province and Pohang City (H.O.). H. O. is sincerely grateful for the KIAS member, and log cabin at POSTECH to provide nice space to come up with this project.}

\end{document}